\documentclass[twocolumn,aps,prd]{revtex4-1}

\usepackage{slashed}
\usepackage{enumitem}

\usepackage{float}

\usepackage{tikz-cd}
	\usepackage{amsmath}
	\usepackage{amsfonts}
	\usepackage{amssymb}
	\usepackage{graphicx}
	\usepackage{mathtools}
	\usepackage{stmaryrd}
	\usepackage{autobreak} 
    \usepackage{cancel}
 
	\allowdisplaybreaks
	\usepackage[colorlinks=true,citecolor=blue,linkcolor=red]{hyperref}
	\usepackage{bbold}					
	\usepackage{multirow}				
	\usepackage[normalem]{ulem}        
	\usepackage{array} 
    \usepackage{cancel} 
\usepackage{lipsum}
\usepackage{graphicx}

\usepackage{titletoc} 
\usepackage{mathtools}

	\newcolumntype{x}[1]{>{\centering\let\newline\\\arraybackslash\hspace{0pt}}p{#1}}
	
	\makeatletter
\newcommand*\rel@kern[1]{\kern#1\dimexpr\macc@kerna}
\newcommand*\widebar[1]{%
  \begingroup
  \def\mathaccent##1##2{%
    \rel@kern{0.8}%
    \overline{\rel@kern{-0.8}\macc@nucleus\rel@kern{0.2}}%
    \rel@kern{-0.2}%
  }%
  \macc@depth\@ne
  \let\math@bgroup\@empty \let\math@egroup\macc@set@skewchar
  \mathsurround\z@ \frozen@everymath{\mathgroup\macc@group\relax}%
  \macc@set@skewchar\relax
  \let\mathaccentV\macc@nested@a
  \macc@nested@a\relax111{#1}%
  \endgroup
}
\makeatother

	\DeclareMathAlphabet{\mathbbold}{U}{bbold}{m}{n}

	\newcounter{subeqn} %
	\makeatletter
	\@addtoreset{subeqn}{equation}
	\makeatother

\definecolor{ZM}{rgb}{.5,0,.5}
\definecolor{SD}{rgb}{0,1,0}

\newcommand\trick[1]{} 

\begin{document}
\title{Interaction-induced topological phase transition at finite temperature}

\author{Ze-Min Huang$^{1}$}

\author{Sebastian Diehl$^1$}

\affiliation{$^1$Institute for Theoretical Physics, University of Cologne, 50937 Cologne, Germany}

\begin{abstract}
We demonstrate the existence of topological phase transitions in interacting, symmetry-protected quantum matter at finite temperatures. Using a combined numerical and analytical approach, we study a one-dimensional Su-Schrieffer-Heeger model with added Hubbard interactions, where no thermodynamic phase transition occurs at finite temperatures. The transition is signalled by a quantized, non-local bulk topological order parameter. It is driven by defects, which are enabled by the combination of interaction and thermal activation, with no counterpart in the non-interacting limit. The defects localize topological zero modes, which, when sufficiently abundant, cause the order parameter to vanish. This phenomenon, interpreted via bulk-boundary correspondence, reflects the loss of a topological edge mode at a well-defined critical temperature in the thermodynamic limit. Unlike zero-temperature topological transitions, these finite-temperature transitions lack thermodynamic signatures but remain observable in controlled quantum systems, such as ultracold fermionic atoms in optical lattices.
\end{abstract}

\maketitle

\textcolor{red}{\textit{Introduction.--}} According to Peierls' argument, there are no thermodynamic phase transitions -- witnessed by local order parameters -- in generic short-range interacting systems in one dimension and at finite temperature \cite{peierls1936mpcps, landau1980pergamon, goldenfeld1992westview}. Consider, for example, an Ising model with local magnetization order parameter. The energy cost for creating a localized domain wall defects is intensive, $E \sim N^0$, where $N$ is the number of spins, but the configurational entropy gain for placing the domain wall $S \sim T \log N$, such that the free energy $F = E - TS$ will be minimized by creating such defects and destroying the local order, at any finite temperature $T>0$.  

\begin{figure}[h!]
\centering
\includegraphics[scale=0.25]{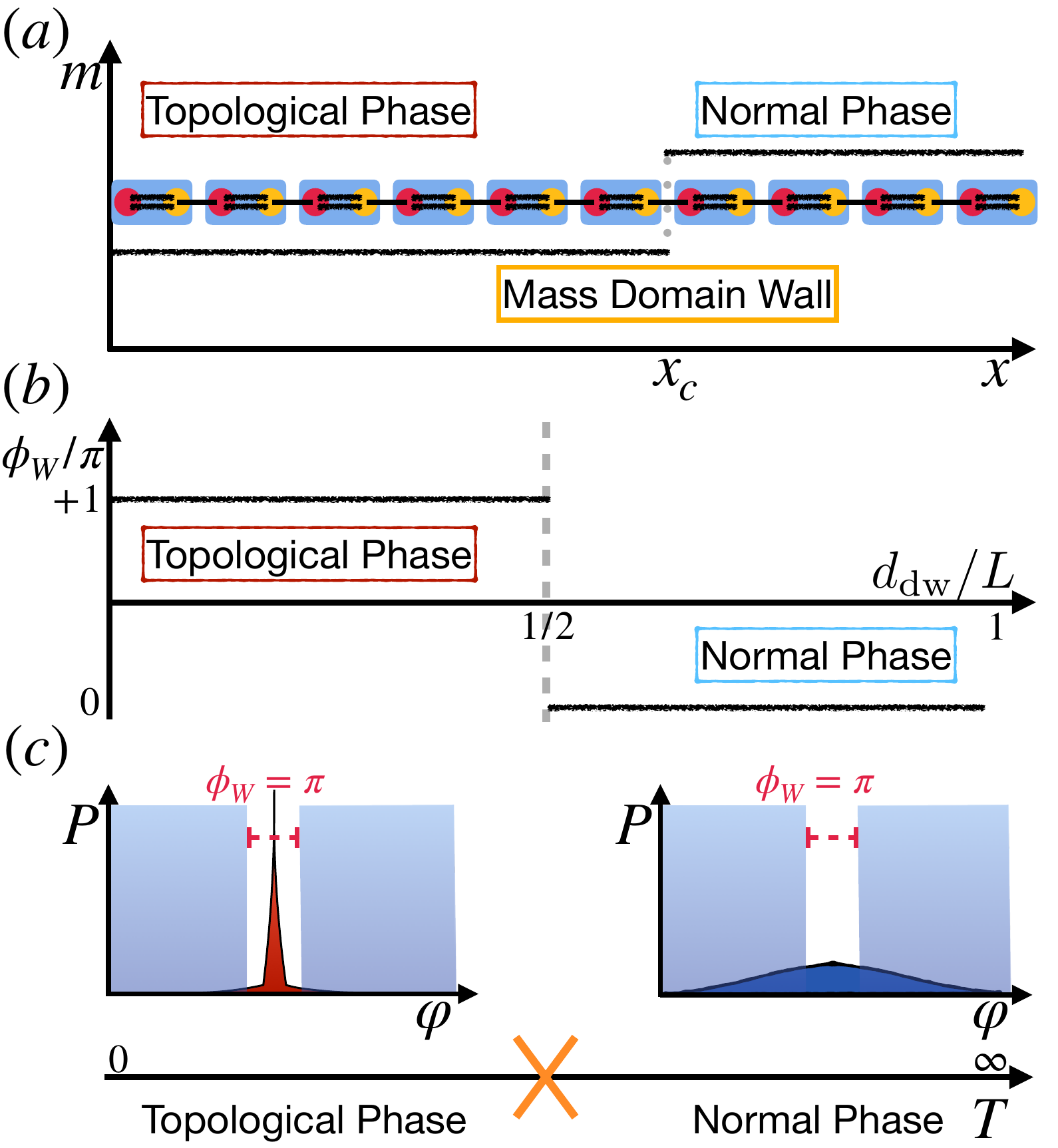}
\caption{Mechanism for the interaction-induced finite temperature topological phase transition. (a,b) Domain wall configurations can counteract a nontrivial topological signal $\phi_W$, Eq.~\eqref{eq:msto}; in the SSH-Hubbard model, mass domain walls are created by the interplay of interaction and thermal activation at finite temperature. (a) visualizes a sharp mass domain wall at $x_c$, separating the topological from a normal segment.  (b) depicts the topological signal as a function of $d_{\text{dw}}$ (the length of the normal phase domain), exhibiting a transition from topological to normal as $d_{\text{dw}}/L$ exceeds a specific threshold. (c)  The effect of interactions $U$ can be described by a fluctuating scalar field $\varphi$, via a Hubbard-Stratonovich decoupling.
It is distributed according to a probability $P[\varphi]$; the topological signal $\phi_W=\pi$ is contributed from a narrow  window for $P$ only (see text for an explanation), while the normal phase signal $\phi_W=0$ occurs outside this region (blue shaded area).
In the limit of zero temperature,  $P$ converges to a Dirac-Delta function, centered at homogeneous $\varphi$, and the signal is lost by the peak moving out of the `topological window'. At high temperatures instead, the problem becomes strictly local, with distribution $P[\varphi] \sim e^{-\frac{\beta}{2U}\sum_i(\varphi_i - m)^2}$, which becomes broad as the inverse temperature $\beta \to 0$ ($m$ is the bare mass). This increases the likelihood for the formation of non-topological segments in the wire.  The interplay of interaction $U$ and thermal fluctuations thus favor the normal phase, indicating an interaction driven topological phase transition at finite temperature due to the proliferation of domain walls.
\label{fig1:conceptual_plot}}
\end{figure}

In this Letter, we show that topological phase transitions -- detected by non-local order parameters -- can exhibit a different pattern, and report the first instance of an interaction-induced topological phase transition in symmetry protected quantum matter at non-zero, finite temperature. This is exemplified for the Su-Schrieffer-Heeger (SSH) chain in the presence of a repulsive Hubbard interaction, whose continuum limit is the Gross-Neveu model \cite{gross1974prd, fradkin2021princeton}, by both analytical and numerical analyses. Again, domain wall defects play a crucial role. They arise due to the simultaneous presence of interactions and finite temperature \cite{dashen1975prd}, and separate topologically non-trivial and trivial regions, see Fig.~\ref{fig1:conceptual_plot}. They thus localize topological zero modes, which causes destructive interference in the non-local order parameter signal. The topological signal is lost only beyond a finite critical temperature, where an extensive fraction of the system has become topologically trivial. This threshold behavior is rooted in the non-local nature of the topological order parameter, this way circumventing Peierls' argument. The transition finds an interpretation as the loss of a topologically protected edge mode
at a temperature which becomes sharply defined in the thermodynamic limit. 

\textcolor{red}{\textit{Mixed state topological phases and their order parameter.--
}} 
To capture symmetry-protected topological order for fermion systems \cite{hassan2010rmp, qi2011rmp,pollmann2010prb,chiu2016rmp, chen2013prb, shizoki2018prb} in mixed states, we consider the expectation value of a global unitary operator chosen based on symmetry principles \cite{hassan2010rmp,qi2011rmp,shizoki2018prb, huang2024arxiv}. This expectation value takes the form of a non-unitary partition function, whose  
phase acts as an order parameter: The phase signal is experimentally observable \cite{bardyn2018prx, huang2024arxiv}, quantized, and cannot change without breaking the protecting symmetry or encountering a zero in the partition function amplitude, akin to a Lee-Yang zero. Thus, two systems are in the same phase if they can be smoothly connected while preserving symmetry and a non-zero partition function; distinct phases are separated by Lee-Yang zeros. 

Specifically, in one dimension and at finite temperature, we consider a unitary operator with support on the entire system \cite{resta1998prl, bardyn2018prx, huang2024arxiv}, which reduces to the Zak phase in the pure state limit \cite{zak1989prl, di2009rmp},  
\begin{equation}
\langle\hat{T}_{X}\rangle\equiv \mathcal{N}^{-1}\text{Tr}\left(\hat\rho\hat{T}_{X}\right),\ \begin{cases}
\hat{T}_{X}\equiv e^{-i\frac{2\pi}{L}\hat{X}}\\
\hat{X}=\sum_{i}x_{i}\hat{n}_i
\end{cases}.\label{eq:Tx}
\end{equation}
$\hat{\rho}$ is the (unnormalized) density matrix for a system with length $L$, $\mathcal{N} = \text{Tr}\hat\rho$.  $\hat{n}_i \equiv\sum_a \frac{1}{2}[\hat{\psi}_{i,\ a}^{\dagger},\ \hat{\psi}_{i,\ a}]$ is the suitably symmetrized local charge density operator, and $\hat{X}$ represents the global position operator with respect to the $x_{i}$-coordinate: $\hat{\psi}_{i,\ a}$ ($\hat{\psi}_{i,\ a}^{\dagger}$)
denotes the fermion annihilation (creation) operator, with the subscript
$i,\ a$ for lattice site and internal index, respectively. We will be interested in symmetry protected matter (e.g. with particle-hole symmetry), in such a way that a reality condition is implemented,  $\langle\hat{T}_{X}\rangle \in \mathbb{R}$. This renders a robust and quantized topological order parameter \cite{qi2011rmp, huang2022prb, tasaki2023jmp,molignini2023prr,xiao2023quantum, huang2024arxiv,mao2024rpp}, 
\begin{equation}
\frac{\phi_W}{\pi}\equiv\frac{1}{\pi}\text{Im}\ln\langle\hat{T}_{X}\rangle=\begin{cases}
1\mod 2\mathbb{Z}, & \text{topological}\\
0\mod 2\mathbb{Z}, & \text{normal}
\end{cases},\label{eq:msto}
\end{equation}
distinguishing topological from normal phases. The quantization is built upon the non-local nature of the order parameter and symmetries, irrespective to considering pure or mixed states. In particular, in translation invariant Gaussian systems, the order parameter will take its zero temperature value for any finite temperature. It can jump from the non-trivial to the trivial value only at infinite temperature \cite{bardyn2018prx, wawer2021prbchern, huang2022prb,pi2022prb, huang2024arxiv, mao2024arxiv}.

\textcolor{red}{\textit{Topological phase transition mechanism.--}} In the following, we will describe how interactions can modify this picture. The key difference to a non-interacting situation lies in the existence of domain walls, which are induced by interactions at any non-zero temperature. Hence, we will first consider the effect of domain walls on the non-local order parameter, and only then move to a concrete model, which is known to host such defects. We will show that $\hat{T}_X$ imparts a phase factor to domain walls due to a topological zero mode localized there. This leads to destructive interference, and enables Lee-Yang zeros of $\langle\hat{T}_X\rangle$, in turn allowing the topological signal to jump. This, however, only takes place at some finite critical temperature -- neither at $T=0$, as in Peierls' thermodynamic transition scenario, nor at $T^{-1}=0$, as in topological transitions in Gaussian states.

Consider first a single, \textit{sharp} domain wall centered at $x_{c}$, formed within a topological insulator of length $L$ ($L\gg d_{\text{lat}}$, $d_{\text{lat}}$ the lattice spacing, see Fig. \ref{fig1:conceptual_plot}
(a)). It demarcates a topological phase on the left from a normal phase on the right \cite{FN_dw}. The bulk-boundary correspondence principle necessitates a zero mode localized at $x_{c}$ and created
by the operator $\hat{\psi}_{x_{c}}^{\dagger}$. This zero mode contributes in two significant ways: First, it ensures the presence of at least twofold degenerate eigenstates of $\hat{\rho}$, denoted by $\left\{ |\psi_{n}\rangle,\ \hat{\psi}_{x_{c}}^{\dagger}|\psi_{n}\rangle\right\} $ with $\hat{\psi}_{x_c}|\psi_n\rangle=0$. Second, it assigns different phases to these states via $\hat{T}_{X}$, i.e., $\left\{ e^{i\frac{\pi}{L}x_{c}},\ e^{-i\frac{\pi}{L}x_{c}}\right\} $, due to the commutation relation $\hat{T}_{X}\hat{\psi}_{x_{c}}^{\dagger}=e^{-i\frac{2\pi}{L}x_{c}}\hat{\psi}_{x_{c}}^{\dagger}\hat{T}_{X}$.
Hence, we can express the order parameter as
\begin{equation}
\langle\hat{T}_{X}\rangle = 2\cos\left(\frac{\pi}{L}x_{c}\right)\times\mathcal{N}^{-1}\text{Tr}_{c^\prime}\left(\hat\rho \hat{T}_X\right),\label{eq:T_X}
\end{equation}
where $\text{Tr}_{c^\prime}$ runs over the complementary Hilbert space with respect to the zero mode. Equation \eqref{eq:T_X} confirms the presence of a phase factor associated with the domain wall centered at $x_c$,  $\cos\left(\frac{\pi}{L}x_{c}\right)$. 

For the choice $x_c=L/2$, we formally encounter a Lee-Yang zero of the  signal. However, this is not yet a physical finding, as it discards the translation invariance associated with the formation of a domain wall. Consider thus two sufficiently spaced domain walls located at $x_c$ and $x_c+d_{\text{dw}}$, respectively, where $d_{\text{dw}}$ denotes the length of the normal phase domain. Owing to zero modes localized within the domain walls, $\hat{T}_X$ imparts to this configuration a factor $\cos(\frac{\pi}{L}x_c)\times\cos[\frac{\pi}{L}(x_c+d_{\text{dw}} )]$ (see Eq. \eqref{eq:T_X}). We then estimate $\langle\hat{T}_X\rangle$ by integrating over all domain-wall configurations connected by a translation, indexed by $x_c$, under the assumption that $\mathcal{N}^{-1}\text{Tr}_{c^\prime}(\hat\rho\hat{T}_X)$ remains largely independent of $x_c$.  This integration introduces an order-one factor $\cos(\frac{\pi}{L}d_{\text{dw}})$ to $\langle \hat{T}_X\rangle$. It indicates a loss of the topological signal when the normal phase domain dominates ($\frac{d_{\text{dw}}}{L}>\frac{1}{2}$). This suggests a threshold behavior: the topological signal is lost only when an extensive fraction of the system is in the normal state. As per the above discussion, this behavior is enabled by the non-local nature of the topological order parameter. This reconciles with Peierls' argument, which builds on local order parameters to exclude finite temperature  thermodynamic phase transitions.

\textcolor{red}{\textit{Finite temperature topological phase transition in the SSH-Hubbard model.--}} Domain walls are ubiquitous in one dimensional interacting systems. One case in point is the Gross-Neveu model, which hosts mass domain wall defects suppressing thermodynamic phase transitions at finite temperature \cite{dashen1975prd}. Here we investigate a lattice variant of this model, to reveal how these defects impact on the integrity of the topological phase: We study $\hat H = \hat H_{\text{SSH}}+ \hat H_{\text{int}}$, with
\begin{align*}
\hat{H}_{\text{SSH}}\left[m\right]&=\sum_{i}\left[\left(\tfrac{1}{2}\hat{\Psi}_{i+1}^{\dagger}(\sigma^{z}-i\sigma^{x})\hat{\Psi}_{i}+\text{h.\ c.}\right)+ m\hat{\Psi}_{i}^{\dagger}\sigma^{z}\hat{\Psi}_{i}\right],\\
\hat{H}_{\text{int}}&=U \sum_{i}\left(\hat{n}_{+,\ i}-\tfrac{1}{2}\right)\left(\hat{n}_{-,\ i}-\tfrac{1}{2}\right), 
\end{align*}
where $\sigma^\ell, \ell= x,y,z$ are the Pauli matrices, the hopping
constant is set to $1$,  $\hat{\Psi}_{i}=(\hat{\psi}_{0,\ i},\ \hat{\psi}_{1,\ i})^{T}$
 a two-component fermion annihilation operator, and $\hat{n}_{\pm,\ i}\equiv\hat{\Psi}^{\dagger}_i\frac{1\pm\sigma^{z}}{2}\hat{\Psi}_i$. $\hat{H}_{\text{SSH}}[m]$ represents the simplified SSH chain containing only one parameter, the mass $m$. It is in its topological (normal) phase for $\text{\ensuremath{\left|m\right|}}<1$
$\left(\left|m\right|>1\right)$.
The additional Hubbard interaction is repulsive, $U>0$, and can be recast
as a Gross-Neveu interaction, $\hat{H}_{\text{int}}=-\frac{U}{2}\sum_{i}\left(\hat{\Psi}_{i}^{\dagger}\sigma^{z}\hat{\Psi}_{i}\right)^{2}+\frac{U}{4}$.
The model is equipped with a  particle-hole symmetry $\hat{\Psi}\rightarrow\sigma^{x}\hat{\Psi}^{\dagger}$,
which not only guarantees a quantized topological order parameter, but also safeguards
the model against the Monte-Carlo sign problem \cite{li2019arcmp}. 

\begin{figure}[t!]
\centering
\includegraphics[scale=0.25]{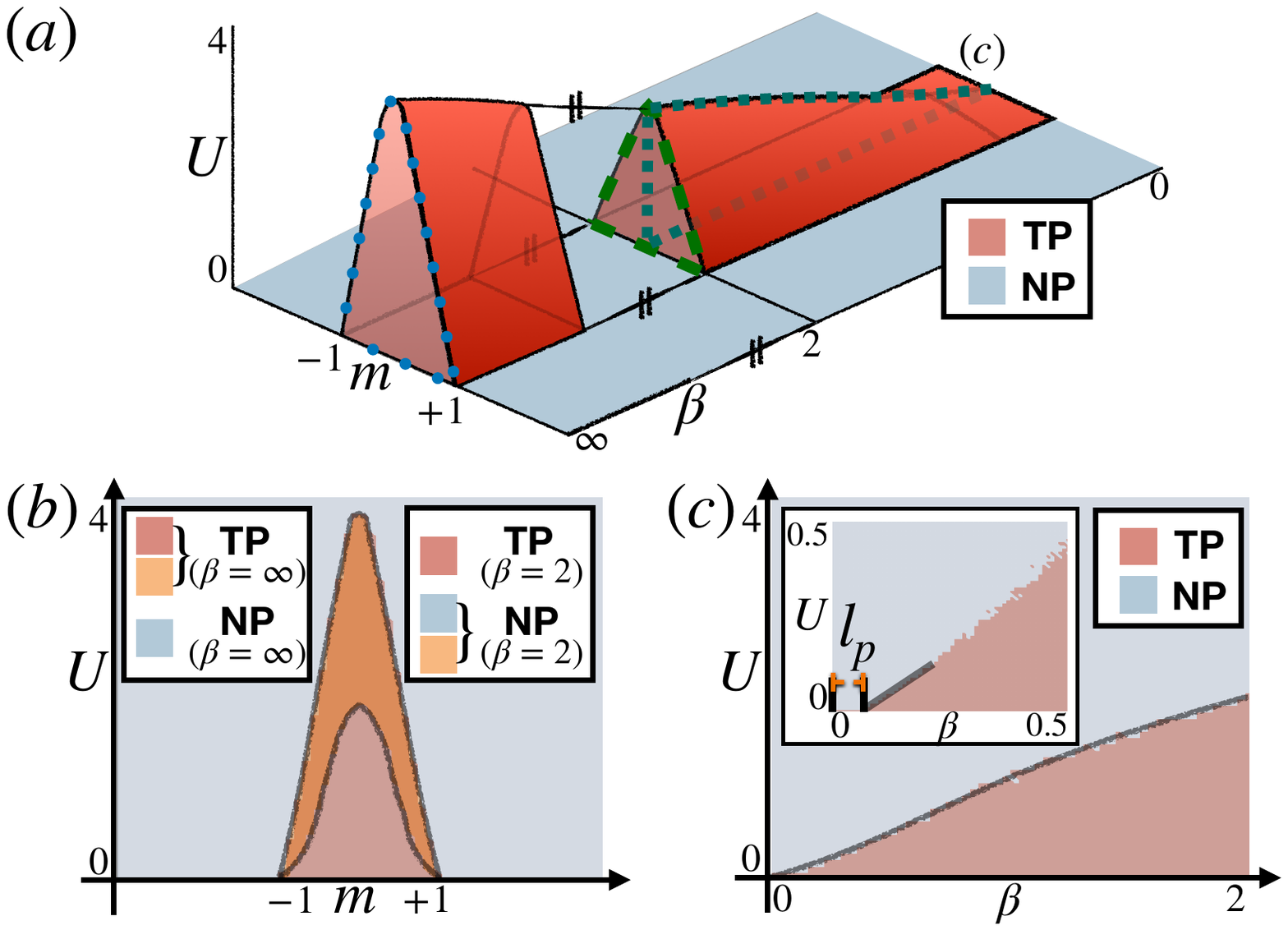}\caption{(a) Schematic topological phase diagram constructed
from DMRG (blue dotted line), and DQMC (green dashed line). The topological order
parameter, Eq. \eqref{eq:msto}, equals $1$
for the topological phase below the red colored dome (denoted by TP, colored red in (b,c)), and $0$
for the normal phase (denoted by NP, colored blue  in (b,c)). Actual data points for the critical lines in different cuts are plotted  in  (b,c) for a $32$-site system: (b) $m-U$ plane at inverse temperatures $\beta=\infty$ and $\beta=2$, (c) $U-\beta$ plane at $m=0$. The inset demonstrates the linear behavior $\beta_c \propto U$ (Eq.~\eqref{eq:aymptT}) of the inverse critical temperature. The offset scales to zero in the thermodynamic limit, cf. \cite{supp}.  \label{fig2:global_phase_diagram}}
\end{figure}

We have constructed the phase diagram of the model by evaluating $\langle \hat{T}_X\rangle$ using complementary density matrix renormalization group (DMRG) and determinant quantum Monte Carlo (DQMC) techniques at zero and finite temperatures, respectively. The resulting topological phase diagram is reported in Fig.~\ref{fig2:global_phase_diagram}, showing good convergence in all temperature regimes for a system with 32 sites and periodic boundary conditions (cf. \cite{supp} for a finite size scaling analysis). 
 To gain analytical insight into the topological phase transition, we utilize a Hubbard-Stratonovich transformation to represent $\langle\hat{T}_{X}\rangle$ as an ensemble average over a fermionic Gaussian state. Here, the four-fermion interaction 
is traded for a scalar auxiliary field $\varphi$, which encompasses
inhomogeneous configurations, including domain walls. Specifically, we obtain
\begin{equation}
\langle\hat{T}_{X}\rangle=\int\mathcal{D}\varphi\;P\left[\varphi\right]\ T_{X}\left[\varphi\right].\label{eq:EGP_prob_dis}
\end{equation}
$P\left[\varphi\right]$ represents the probability distribution for the $\varphi$
field (see Fig. \ref{fig1:conceptual_plot} (c)), given as 
\begin{equation}
P\left[\varphi\right]=\mathcal{N}^{' -1} e^{-\frac{1}{2U}\int_{0}^{\beta}d\tau\sum_{i}\left(\varphi_{i}-m\right)^{2}}\det\left(\partial_{\tau}+H_{\text{SSH}}\left[\varphi\right]\right).\label{eq:P=00005Bphi=00005D}
\end{equation}
Here we absorbed the $m$-dependence of the fermion determinant into the Gaussian prefactor, so $H_{\text{SSH}}[\varphi]$ renders a topological signal for $|\varphi|<1$. $P$ remains positive due to the absence of the sign problem; $\mathcal{N}'$ is a normalization factor.
$\tau$ represents the imaginary time associated with the inverse temperature $\beta=1/T$. The  factors in $P\left[\varphi\right]$ describe competing effects: The Gaussian component tends to pin $\varphi$ at $m$. The fermion determinant instead,  with $H_{\text{SSH}}\left[\varphi\right]$ representing the first quantized SSH Hamiltonian in the presence of the $\varphi$ field, favors large values of $\varphi$ \cite{fn_ph2}, competing with the Gaussian term. Finally, $T_{X}\left[\varphi\right]$ is the expectation value of $\hat{T}_{X}\left[\varphi\right]$ within the non-interacting SSH Gibbs ensemble, 
\begin{equation}
T_{X}\left[\varphi\right]=\text{Tr}\left(\frac{\hat{\rho}_{\text{SSH}}\left[\varphi\right]}{\text{Tr}\hat{\rho}_{\text{SSH}}\left[\varphi\right]}\hat{T}_{X}\right),\ \hat{\rho}_{\text{SSH}}\left[\varphi\right]\equiv\mathcal{T}_\tau e^{-\int_{0}^{\beta}d\tau\hat{H}_{\text{SSH}}\left[\varphi\right]},
\end{equation}
where $\mathcal{T}_\tau$ stand for the time-ordering operator, originating from the imaginary
time dependent $\varphi$-field. 

Analyzing $P[\varphi]$ (see Eq. (\ref{eq:EGP_prob_dis})) in different temperature regimes provides insights into the phase diagram shown in Fig. \ref{fig2:global_phase_diagram}. In the zero-temperature limit, the triangular shape of the zero temperature critical line (cf.  Fig. \ref{fig2:global_phase_diagram}
(b) and  \cite{guo2011prb, juneman2017prx,rachel2018rpp}) is reproduced in a self-consistent mean field treatment. The dominant effect of interactions is a Hartree shift of the mass term  \cite{guo2011prb, juneman2017prx,rachel2018rpp},
because $P[\varphi]$ favors a homogeneous $\varphi$-field configuration \cite{fradkin2021princeton, altland2010oxford}. At low temperatures $\beta \to \infty$, this leads to a sharply
peaked structure of $P[\varphi]$ at a specific $\varphi$-field configuration (see Fig.~\ref{fig1:conceptual_plot} (c)). The transition then occurs when this peak shifts out of the `topological window' $\left|\varphi\right|<1$ in an effective single-particle picture.  

Conversely, in the high-temperature regime, $P[\varphi]$ broadens significantly, leading to the proliferation of domain walls and thus a trivial topological signal, as we argue now. To this end, we compute $P\left[\varphi\right]$ systematically in a high-temperature expansion in the thermodynamic limit. 
Specifically, the fermion determinant term in Eq.~\eqref{eq:P=00005Bphi=00005D} to leading order in $\beta\ll 1$ (in units of the hopping constant) is given by
\begin{equation}
\ln\det\left(\partial_{\tau}+H_{\text{SSH}}\left[\varphi\right]\right)=\tfrac{1}{8}\beta^{2}\text{Tr}\left(H_{\text{SSH}}\left[\varphi\right]\right)^{2}= \tfrac{1}{4}\beta^2\sum_i\varphi_i^{2},\label{eq:P_highT}
\end{equation}
up to an irrelevant constant independent of the $\varphi$ field. The expansion exclusively relies on the smallness of the imaginary time extent $\beta$, and the second equality is a direct consequence of tracing over the tight-binding SSH chain Hamiltonian. Importantly, it is thus applicable to a spatially inhomogeneous $\varphi$ field configurations, but turns out to be quadratic in $\varphi_i$, and strictly local in this limit.

We are now positioned to explore the critical temperature by estimating the length of topological domains. Given the strict locality found in  the high-temperature limit, we focus on the single-site probability $p$ of $\varphi_i$ to be within the topological window ($\left|\varphi\right|<1$). This is given by 
\begin{equation}
p=\mathcal{N}_{p}^{-1}\int_{-1}^{1}d\varphi_i e^{-\frac{1}{2U}\beta\left(\varphi_i-m\right)^{2}}\times e^{\frac{1}{4}\beta^{2}\varphi_i^{2}},\label{eq:prob_tp}
\end{equation}
where the integration boundaries restrict to integrating over configurations in the topological window, whereas the integration in the normalization factor $\mathcal{N}_{p}=\int_{-\infty}^{+\infty}d\varphi_i e^{-\frac{1}{2U}\beta\left(\varphi_i-m\right)^{2}}\times e^{\frac{1}{4}\beta^{2}\varphi_i^{2}}$ is unconstrained. This formulation elucidates the primary impact of temperature elevation: the broadening of the
probability distribution. Given the large parameter space for the normal
phase ($\left|\varphi\right|>1$) compared to the topological phase ($\left|\varphi\right|<1$), thermal fluctuations favor the normal phase and counteract the topological signal. Specifically, for a system of length $L$ with $N=L/d_{\text{lat}}$ sites, the probability that $|\varphi|<1$ within $n$ sites follows a binomial distribution $\frac{N!}{(N-n)!n!} p^n(1-p)^{N-n}$. The mean of this distribution, representing the expected length of topological domains, is $L p$. The critical (inverse) temperature is estimated via the point where the domains of the topological and normal phases are comparable, i.e., $\frac{d_{\text{dw}}}{L}=1-p\sim\frac{1}{2}$, with $d_{\text{dw}}$ the extent of the non-topological segment. Finally, considering Eq.~\eqref{eq:prob_tp} to the leading order of $\beta$  (i.e., $p=\sqrt{\frac{2}{\pi}}\sqrt{\frac{\beta}{U}} +\mathcal{O}(\beta^{3/2})$), we can deduce the scaling of the critical (inverse) temperature 
\begin{equation}\label{eq:aymptT}
\beta_{c}\propto U.
\end{equation}
As interactions vanish, $U\to 0$, and domain walls can no longer be created, the critical temperature diverges, in line with the behavior of non-interacting systems. More quantitatively, the asymptotic scaling of the critical temperature aligns with the one obtained numerically, cf. Fig.~\ref{fig2:global_phase_diagram} (c).

\begin{figure}
\includegraphics[scale=0.2]{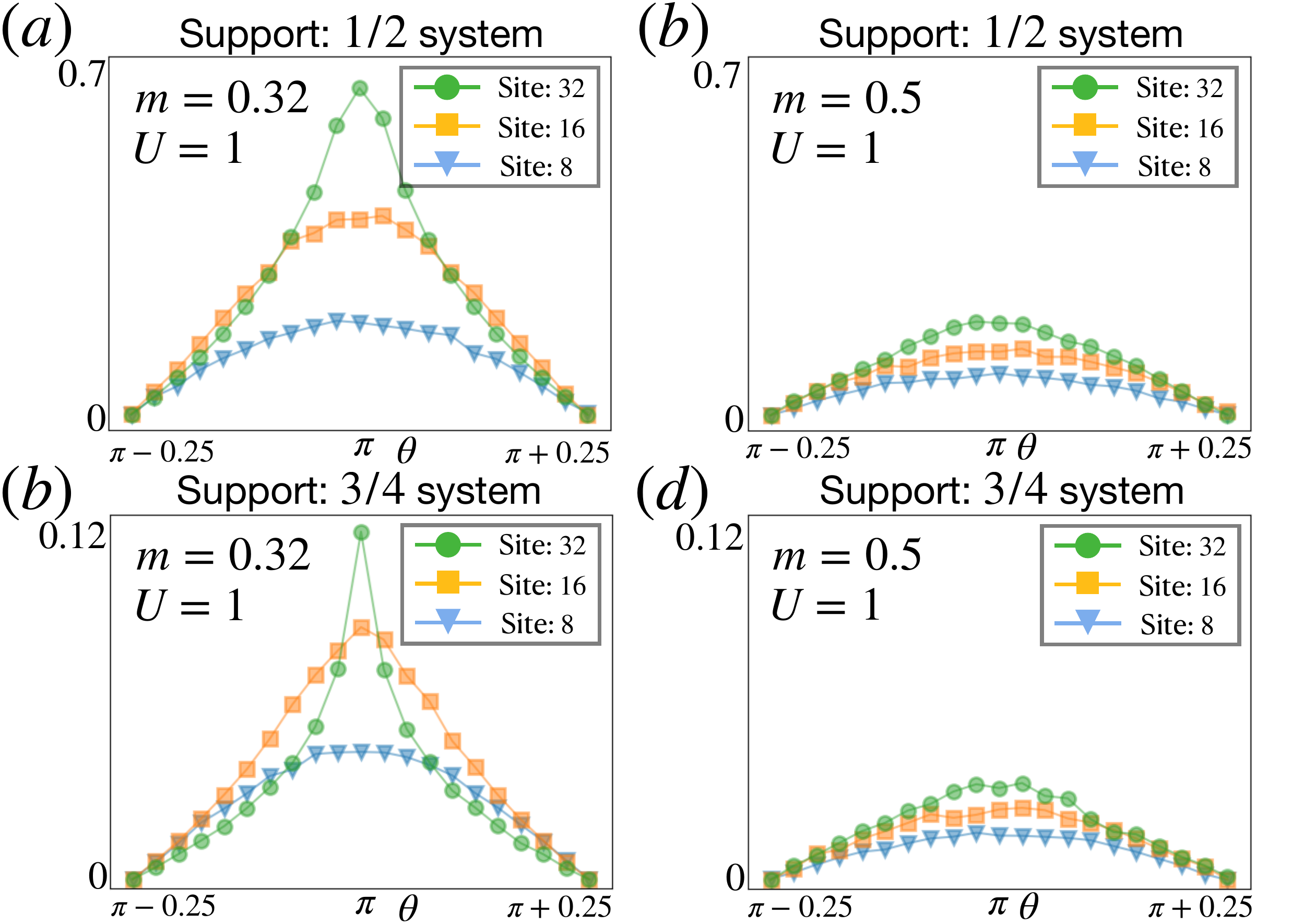}
\caption{Boundary fermion parity in topological  (a, c) and  normal  (b, d) regimes for different system sizes (i.e., $N=8, 16, 32$) in an SSH-Hubbard chain with open boundary conditions at $\beta=2$. $\hat Q_{\text{bd}}$ is supported over $1/2$ (a, b) and $3/4$ (c, d) of the system size, to minimize the finite-size effect: A necessary condition for bulk-boundary correspondence is that the boundary fermion parity operator have a support larger than the boundary localization length \cite{supp}. We plot $-\log_{10}|\langle e^{-i \theta \hat{Q}_{\text{bd}}} \rangle_{\text{OBC}}|/N$ (with minimum across all angles $\theta$ subtracted). In (a), we identify a cusp at $\theta=\pi$, which sharpens and rises as the system size increases, indicating a zero of the boundary parity in the thermodynamic limit. In the normal regime (b) instead, no such cusps appear.
\label{fig3:finite_size}}
\end{figure}

\textcolor{red}{\textit{Bulk-boundary correspondence and finite-size effects.--}}  The finite temperature topological phase transition affords a physical interpretation as the loss of a topological edge mode, which is sharply defined in the thermodynamic limit.
To this end, we consider a system with open boundary conditions (OBCs), and study the boundary fermion parity $\langle (-)^{\hat{Q}_{\text{bd}}}\rangle_{\text{OBC}}$, where $\hat{Q}_{\text{bd}}$ is the charge operator with support in the region near the boundary. We demonstrate numerically that this quantity vanishes in the topological phase, and remains finite otherwise, in the thermodynamic limit; conversely, in a finite system, it remains finite everywhere in parameter space. In practice, this vanishing is revealed through cusps of increasing sharpness in the exponent of $|\langle e^{-i\theta\hat{Q}_{\text{bd}}}\rangle_{\text{OBC}}|$ at $\theta=\pi$, which develop in the thermodynamic limit within the topological phase, cf. Fig.~\ref{fig3:finite_size}. 
The interpretation of this behavior proceeds via the bulk-boundary correspondence: A non-trivial bulk signal $\phi_W $ implies the vanishing of the boundary fermion parity, and in turn the existence of anomalous boundary states. This results from the combined particle-hole and \textit{large} $U(1)$ symmetries. It is well-appreciated in interacting systems in their ground state \cite{hassan2010rmp, qi2011rmp, tasaki2023jmp, wen2013prd}, and can be generalized to mixed states via anomaly matching (rather than via perturbative anomaly inflow) \cite{callan1985npb,elitzer1986npb,gaiotto2017jhep, tong2018gauge}. For a quantitative assessment of the bulk-boundary correspondence and additional numerical results, see \cite{supp}.
The thermodynamic limit is crucial as it prevents the hybridization between boundary states at opposite ends, preserving their degenerate nature indicated by a vanishing boundary fermion parity. 

\textcolor{red}{\textit{Observable consequences.--}} The finite temperature topological phase transitions are not witnessed in local thermodynamic observables, like local order parameters and the specific heat, or by divergent length and time scales -- this is ruled out by Peierls' argument. Yet, the global topological order parameter characterizing it is experimentally observable, e.g., via interferometry or full counting statistics in cold atomic systems \cite{bakr2009nature,sherson2010nature,bardyn2018prx, jiang2023scipost, wang2023arxiv,zang2023arxiv, huang2024arxiv, mao2024arxiv}. 
At zero temperature instead, the topological phase transition is accompanied by a spectral gap closing, and thus a thermodynamic transition \cite{resta1999prl, souza2000prb, tasaki2023jmp}.

\textcolor{red}{\textit{Conclusion.--}} Topological phases in mixed states can be captured via a experimentally observable non-local order parameters, which describe topological terms in the partition function and reflect anomalous boundary states. Phase transitions are marked by Lee-Yang zeros, representing a mixed-state generalization of the gap closing condition and providing a finer structure in the phase diagram compared to previous studies based on quantum circuits \cite{roberts2017pra}. We have identified a novel topological phase transition caused by the proliferation of domain walls, resulting from the combined effect of interaction and thermal fluctuations. Central to our findings is the non-local nature of the mixed state topological order parameter, circumventing Peierls' argument and serving as a bulk indicator of the bulk-boundary correspondence. It attaches a phase factor to domain walls. This bears similarities to the phase transition in disordered topological matter at zero temperature \cite{li2009prl,fu2012prl, konig2012prb, altland2014prl}, in that both are caused by decorated defects, but they are distinct by the origin of the attached phase factor. The mechanism seems general, suggesting such transitions to occur in other instances of symmetry protected quantum matter and higher dimensions, beyond the SSH-Hubbard model studied here. The phenomenology of the mixed state transition, which proceeds without standard thermodynamic (local) signatures, motivates exploring the connection to transitions in the quantum information content of mixed states, which come with a similar phenomenology \cite{fan2024prxQ}. More generally, such transitions present new avenues for many-body physics, by exploring the relationship between mixed state topology and topological quantum memory under noise \cite{dennis2002jmp,hastings2011prl, huang2024ci, fan2024prxQ}.

\begin{acknowledgments}
We thank  Alex Altland, Michael Buchhold, Bo Han,  Johannes Lang, David Mross, Achim Rosch, Xiao-Qi Sun, and Guo-Yi Zhu for discussions. Z.-M.H. and S.D. are supported by the  Deutsche Forschungsgemeinschaft (DFG, German Research Foundation) under Germany’s Excellence Strategy Cluster of Excellence Matter and Light for Quantum Computing (ML4Q) EXC 2004/1 390534769 and by the DFG Collaborative Research Center (CRC) 183 Project No. 277101999 - project B02.
\end{acknowledgments}

\clearpage 

\begin{center}
\textbf{\large Supplemental Material for "Interaction-induced topological phase transition at finite temperature"}
\end{center}

\setcounter{equation}{0}
\setcounter{figure}{0}
\setcounter{table}{0}
\setcounter{page}{1}
\setcounter{section}{0}
\makeatletter
\renewcommand{\theequation}{S\arabic{equation}}
\renewcommand{\thefigure}{S\arabic{figure}}
\renewcommand{\thesection}{S\arabic{section}}
\renewcommand{\bibnumfmt}[1]{[S#1]}
\renewcommand{\citenumfont}[1]{S#1}

This supplemental material includes details for: (i) Review of mixed-state topological responses; (ii) Analytical and numerical analysis of the bulk-boundary correspondence; (iii) numerical results for the phase diagram in the $U-\beta$ plane for different system sizes.

\section{Review of topological response in mixed state}
We review the definition of topological phases through their physical signatures — specifically, topological terms in the partition function \cite{hassan2010rmp, qi2011rmp, huang2022prb, huang2024arxiv}, and quantized bulk responses manifesting via bulk-boundary correspondence (see Sec. \ref{supp_sec:bb_correspondence} for more details). This symmetry-based approach provides an integrated framework for topological phases at zero \cite{hassan2010rmp, qi2011rmp} or finite temperatures \cite{huang2022prb,huang2024arxiv}.

\textcolor{red}{\textit{Mixed state topological order parameter.--}} For a fermionic Gibbs state $\hat\rho$, the phase factor of a non-local unitary operator (denoted as $e^{-i\hat{W}}$), chosen based on a symmetry principle \cite{huang2024arxiv}, encodes topological information. The associated expectation value is defined as,
\begin{equation} Z_W[w] \equiv \mathcal{N}^{-1} \operatorname{Tr}\left(\hat{\rho} e^{-i \hat{W}[w]}\right), \quad \hat{W} = \sum_i w_i \hat{\psi}^\dagger_{i, a} \mathcal{W}_{ab} \hat{\psi}_{i, b},
\end{equation} 
where $\hat{\psi}_{i, a}$ ($\hat{\psi}^\dagger_{i, a}$) are fermionic annihilation (creation) operators at site $i$ with internal indices $a$, and $w$ is a slowly varying function of the spatial coordinates. The corresponding phase factor is,
\begin{equation} 
S_{\text{eff}}[w] \equiv \text{Im}\ln\left(Z_W[w]\right). \end{equation} 
 $Z_W[w]$ ($S_{\text{eff}}$) can be viewed as the partition function (action) with an imaginary external field introduced via $e^{-i \hat{W}}$. For a $U(1)$-symmetric $\hat\rho$ and $\mathcal{W}=\mathbb{I}$,  $w$ couples to the charge density and represents the temporal Wilson loop of the $U(1)$ gauge field $A_\mu$ \cite{huang2022prb},
\begin{equation}
a_0\equiv \oint A_0 dt = w.
\end{equation} 
$S_{\text{eff}}$ is a topological term (illustrated below using the Dirac model) that detects the underlying topology. For example, in one dimension, choosing $\hat{W} = \sum_i \frac{2\pi}{L} x_i \hat{n}_i$, $S_{\text{eff}}[a_0=\frac{2\pi}{L}x_i]$ extracts the Zak phase (ensemble geometric phase) at zero (finite) temperature \cite{zak1989prl, di2009rmp, bardyn2018prx, huang2022prb, huang2024arxiv}. For general symmetry classes of fermions, $S_{\text{eff}}$ remains a robust indicator of the underlying topology, provided $\hat{W}$ is chosen appropriately based on symmetry criteria; see Refs. \cite{huang2024arxiv, huang2022prb} for further details. Here we focus on the case with $\mathcal{W} = \mathbb{I}$; hence, $w$ is replaced by $a_0$ hereafter, while $Z$ is retained to denote the resulting partition function. 
Finally, given $e^{-i\hat{W}}$ is a unitary operator, it can be measured in controlled quantum systems, e.g., via interferometry or full counting statistics \cite{bardyn2018prx, huang2024arxiv}. 

\textcolor{red}{\textit{Phase transition and Lee-Yang zero.--}} In this framework, changes in topological responses are accompanied by zeros in the amplitude $|Z[a_0]|$, known as Lee-Yang zeros in statistical mechanics. This occurs because formally, the non-local operator $e^{-i \hat{W}}$ introduces an imaginary external field into the partition function. From an alternative perspective of dynamical phase transitions \cite{heyl2018rpp}, $e^{-i \hat{W}}$ induces quenched dynamics, so $Z[a_0]$ represents the Loschmidt echo, with singularities in $|Z[a_0]|$ marking transition points.

Below, we illustrate this setup using the Dirac stationary state, i.e., $\hat\rho = e^{-\beta \hat H}$, where the $\hat H$ is a Dirac Hamiltonian, and $\beta $ the inverse temperature.

\textcolor{red}{\textit{Example: Dirac stationary state in even spatial dimensions.--}} The effective action $S_{\text{eff}}$ for slowly varying $a_0(\boldsymbol{x})$ can be evaluated from the representative Dirac model at finite temperature \cite{huang2022prb, huang2024arxiv}, from which we obtain a quantized signal robust against thermal fluctuations, linking to the anomaly associated to large $U(1)$ transformations. This reveals a clear distinction between perturbative and large anomalies, as the former is sensitive to thermal fluctuations. For simplicity, we focus on even spatial dimensions, deferring the one-dimensional case to Sec. \ref{supp_sec:bb_correspondence}. Specifically, we find (for details, see Refs. \cite{huang2022prb, huang2024arxiv})
\begin{equation}
S_{\text{eff}}=\text{ch}_W \times \int d^{2n}\boldsymbol{x}\ \mathcal{I}_W[a_0(\boldsymbol{x})]\ \mathfrak{C}^{(2n)}(\boldsymbol{x}),
\end{equation}
where $\mathfrak{C}^{(2n)}$ is the Chern character density, 
\begin{equation}
\mathfrak{C}^{(2n)}(\boldsymbol{x})=\frac{\epsilon^{0i_1i_2\dots i_{2n}}}{(2\pi)^n n!}\partial_{i_1}A_{i_2}\dots\partial_{i_{2n-1}}A_{i_{2n}}.
\end{equation}
$\mathcal{I}_W$ is model dependent. For a massive Dirac fermion with mass $m$ at inverse temperature $\beta$, by taking $\mathcal{W}=\mathbb{I}$, we get
\begin{equation}
\mathcal{I}_{W}\equiv\text{Re}\left\{ -2i\ln\left[\cos\left(\frac{a_0}{2}\right)+i\tanh\left(\frac{\beta|m|}{2}\right)\sin\left(\frac{a_0}{2}\right)\right]\right\}, 
\end{equation}
and 
\begin{equation}
\text{ch}_W=\frac{1}{2}[-1 +\text{sign}(m)]\in\mathbb{Z}.
\end{equation}
where $A_\mu$ is taken to be static for simplicity. 

In the zero-temperature limit, this reduces to the Chern-Simons term, 
\begin{equation}
\beta\rightarrow \infty:\ S_{\text{eff}}=\text{ch}_W\int d^{2n}\boldsymbol{x}\int dt \ A_0\ \mathfrak{C}^{(2n)}(\boldsymbol{x}),
\end{equation}
featuring quantized Hall conductance, 
\begin{equation}
\beta\rightarrow \infty:\ j^0=-\sigma_{H}\ \mathfrak{C}^{(2n)}, \ \sigma_H =\text{ch}_W\in\mathbb{Z},
\end{equation}
with $j^0\equiv -\frac{\delta S_{\text{eff}}}{\delta A_0}$ the charge density.
This quantized local response is directly tied to the perturbative anomaly, obtained by varying the action with respect to $A_\mu$.

However, at finite temperature,  the Hall conductance, a local thermodynamic observable, loses its quantization:
\begin{equation}
\text{finite}\ \beta:\ \sigma_{H}=\text{ch}_W \tanh(\frac{\beta|m|}{2}),
\end{equation}
reflecting that the perturbative anomaly, associated with small variation of gauge field, is susceptible to thermal fluctuations \cite{callan1985npb, stone1991aop, huang2022prb}. In contrast, the non-local mixed state order parameter is robust against temperature, 
\begin{equation}
\oint \frac{da_0}{2\pi}\partial_{a_0} S_{\text{eff}}=\frac{1}{2\pi}S_{\text{eff}}|_{a_0}^{a_0+2\pi}=\text{ch}_W \mathfrak{C}^{(2n)}\in \mathbb{Z},
\end{equation}
representing the winding number of the partition function, measurable via interferometry or full-counting statistics. This reflects a non-perturbative anomaly from large $U(1)$ transformations, which is temperature-independent, unlike the perturbative anomaly.

Finally, a phase transition, characterized by a sudden jump in the mixed-state order parameter, occurs when $\beta|m|=0$ and thus $Z[a_0]=0$. This confirms the absence of a finite-temperature transition in the Gaussian states considered here.

For more information on Dirac stationary states in various dimensions and symmetry classes, we refer to Refs. \cite{huang2024arxiv, huang2022prb}.

\section{$\phi_W$ as a bulk indicator of the bulk-boundary correspondence \label{supp_sec:bb_correspondence}}

We concentrate here on the derivation of  the bulk-boundary correspondence for mixed states in the thermodynamic limit. To this end, it is useful to first generalize Eq.~\eqref{eq:Tx} in the main text to 
\begin{equation}
Z[a_0]\equiv \mathcal{N}^{-1}\text{Tr}[\hat{\rho} e^{-i \sum_{i} a_0(x)\hat{n}_{i}}],\ \text{with}\ \mathcal{N}=\text{Tr}\hat\rho,
\end{equation}
reducing to Eq.~\eqref{eq:Tx}  for $a_0(x)  = \frac{2\pi x }{L}$. We will be interested here in the phase factor, $S_{\text{eff}}$, associated with the thus generalized partition function
\begin{equation}
S_{\text{eff}}[a_0]\equiv \text{Im}\ln Z[a_0].
\end{equation}
For notational simplicity, we interchangeably use $x$ and $i$ to denote the spatial coordinate. The operator $\hat{n}_{i}\equiv \sum_a \frac{1}{2}[\hat{\psi}^\dagger_{i, a},\ \hat{\psi}_{i,\ a}]$ denotes the suitably symmetrized density operator at site $i$ associated with an even dimensional internal space, $a\in  2\mathbb{Z}$, whose eigenvalues are integer. The field $a_0(x)$ can be  interpreted as the temporal Wilson loop of the zero component of a $U(1)$ gauge field, $a_0(x)\equiv\oint dt A_0(t,x)$, where the integration is along the closed imaginary time path in the Matsubara formalism \cite{altland2010oxford, deser1997prl, dunne1997prl}, or along the closed Keldysh contour in the real time formalism \cite{kamenev2011cambridge, altland2010oxford, huang2022prb}. 

Symmetries impose crucial constraints: 
\begin{align*}
\text{Particle-hole symmetry:\quad } & Z\left[a_{0}\right]=Z\left[-a_{0}\right]=Z\left[a_{0}\right]^*,\\
\text{Large\ \ensuremath{U\left(1\right)}\ invariance:\quad} & Z\left[a_{0}\right]=Z\left[{a_0+2\pi \zeta }\right], \ \zeta(x)\in\mathbb{Z},
\label{supp_eq:symmetry_action}
\end{align*}
where $\zeta(x)$ is an integer-valued function, making $a_0(x)$ a compact variable.
The first one pins $S_{\text{eff}}$ at $0$ or $\pi$, i.e., $S_{\text{eff}}\in \pi \mathbb{Z}$, hinting at its role as a topological theta term. The second constraint emerges from the integer eigenvalues of $\hat{n}_i$, reflecting an onsite symmetry from shifting $a_0$ at site $x$ by $2\pi$, while keeping it unchanged at other sites. Under periodic boundary conditions, this allows topologically non-trivial configurations for {\textit{continuous}} $a_0(x)$, characterized by an integer winding number,
\begin{equation}
\frac{1}{2\pi}\int dx \partial_x a_0(x) \in \mathbb{Z}.
\end{equation}
Together, these two symmetries enforce a non-perturbative constraint upon $S_{\text{eff}}$,
\begin{equation}
 S_{\text{eff}} [a_0]\in \pi \mathbb{Z}\ \ \text{for}\ \ \frac{1}{2\pi}\int dx \partial_x a_0(x) \in\mathbb{Z}.\label{supp_eq:S_eff_constraint}
\end{equation}

For slowly varying $a_0$, this non-perturbative constraint links $S_{\text{eff}}$ to the topological theta term via gradient expansion (see e.g., \cite{ginsparg1980npb} for finite-temperature effective field theory). Specifically, under periodic boundary conditions and to leading order in spatial derivatives, $S_{\text{eff}}$ is given as,
\begin{equation}
S_{\text{eff}}[a_0] = \pi\times \text{ch}_W[\beta, U, m]\times \left\{\frac{1}{2\pi}\int dx\partial_{x}\mathcal{I}_W[a_0(x)] \right\}, \label{supp_eq:S_eff}
\end{equation}
which encompasses results calculated from the Dirac model as a specific example \cite{huang2022prb, huang2024arxiv}.
Here, $\text{ch}_W[\beta, U, m]\in \mathbb{Z}$ is a function of the model parameters. The function $\mathcal{I}_W[a_0]$ incorporates the variable $a_0$ to all orders, which is essential for maintaining the large $U(1)$ invariance \cite{deser1997prl, dunne1997prl}. While its specific form is model dependent, it features an important property: $\mathcal{I}_W|_{a_0}^{a_0 +2\pi}=2\pi$ \cite{FN_Iw} necessary to support topologically non-trivial configurations of $a_0$. For example, with $\frac{1}{2\pi}\int \partial_x a_0(x)=\frac{1}{2\pi}[a_0(L)- a_0(0)]=1$, it follows that $S_{\text{eff}}\in \pi \mathbb{Z}$ due to $\int dx\partial_x\mathcal{I}_W[a_0(x)]=\mathcal{I}_W|_{a_0(0)}^{a_0(L)}=2\pi$. Moreover, the $\pi \mathbb{Z}$ quantization of $S_{\text{eff}}$ is crucial: it not only prohibits the existence of any function that depends solely on $a_0$ (like $\int dx f_0[a_0]$), but also mandates linear (spatial) gradient terms (e.g., of the form $\int dxf_1[a_0]\partial_{x}a_0$) to manifest as a total derivative. In the zero temperature limit, $S_{\text{eff}}$ aligns with the celebrated topological theta term, manifesting as $\mathcal{I}_W[a_0]= a_0$. 

The action in Eq.~\eqref{supp_eq:S_eff} contains non-perturbative information about boundaries, which is linked to large gauge invariance, highlighting its relevance to mixed states with finite $\beta$. Specifically, in the following subsections, we will focus on the open-boundary scenario, analytically demonstrating (Sec.~\ref{supp_sec:bbc_analytical}) and numerically confirming (Sec.~\ref{supp_sec:bbc_numerical}) the bulk-boundary correspondence.

\subsection{Bulk-boundary correspondence from anomaly matching \label{supp_sec:bbc_analytical}}
Here, we will demonstrate  that an odd $\text{ch}_W$ necessitates a vanishing boundary fermion parity via anomaly matching (see, e.g., \cite{callan1985npb,gaiotto2017jhep, tong2018gauge}),  
\begin{equation}
0=\frac{\langle (-1)^{\hat{Q}_{\text{bd}}}\rangle_{\text{OBC}}}{ |\langle  (-1)^{\hat{Q}}\rangle_{\text{PBC}}|^{\frac{l_{\text{bd}}}{L}}},\label{supp_eq:fp_bd_charge}
\end{equation}
where the subscript $\text{OBC}$ and $\text{PBC}$ emphasize the difference between open and periodic boundary conditions. $\hat{Q}_{\text{bd}}\equiv\sum_{i\in \text{bd},\ a}\frac{1}{2}[\hat{\psi}^\dagger_{i, a}, \hat{\psi}_{i, a}]$ is the symmetrized boundary charge operator, with support in the boundary region denoted by 'bd' of length $l_{\text{bd}}$. The numerator can be represented as $\langle e^{-i\sum_i a_0(x)\hat{n}_i}\rangle_{\text{OBC}}$, with $a_0=\pi$ in the boundary region, and $0$ otherwise. The denominator $|\langle (-1)^{\hat{Q}} \rangle_{\text{PBC}}|^{l_{\text{bd}}/L}$, assumed nonzero, is introduced for conceptual and numerical convenience, rendering $\frac{\langle (-1)^{\hat{Q}{\text{bd}}} \rangle{\text{OBC}}}{|\langle (-1)^{\hat{Q}} \rangle_{\text{PBC}}|^{l_{\text{bd}}/L}}$ an order-one quantity (e.g., for even $\text{ch}_W$). We refer to this phenomenon as the \textit{bulk-boundary correspondence} for the fermion Gibbs state. This result is well appreciated in two limiting cases:  For Gaussian states, a vanishing boundary fermion parity indicates the existence of boundary zero modes \cite{huang2022prb, huang2024arxiv,mao2024arxiv}, a consequence of Wick's theorem. In the zero temperature limit, this condition matches the pure-state bulk-boundary correspondence \cite{hassan2010rmp,qi2011rmp, wen2013prd, tasaki2023jmp}: an odd $\text{ch}_W$ indicates the presence of either gapless or degenerate boundary states, as a non-degenerate boundary state will not exhibit vanishing fermion parity.

We derive this result, Eq.~\eqref{supp_eq:fp_bd_charge}, by analyzing a system subject to open boundary conditions, where the left and right boundaries are located at $x_{L}$ and $x_{R}$, respectively. To this end, we denote the resulting \textit{full} partition function as $Z^{(o)}[a_0]$, where the superscript $(o)$ highlights the use of open boundaries. $Z^{(o)}[a_0]$ maintains both particle-hole and large $U(1)$ symmetry. For comparison, we also introduce $Z^{(p)}[a_0]$, which contains the same terms as the partition function under periodic boundary conditions, indicated by the superscript $(p)$, and thus encodes the \textit{bulk} information. $Z^{(p)}[a_0]$ preserves particle-hole symmetry, but it may lack large $U(1)$ invariance due to an anomaly (\cite{gaiotto2017jhep,tong2018gauge}, and see below). The \textit{boundary} effect can be quantified by considering the ratio,
\begin{equation}
Z_{\text{bd}}[a_0] \equiv Z^{(o)}[a_0]/Z^{(p)}[a_0],\label{supp_eq:boundary_Z}
\end{equation}
which can deviate from unity due to the boundary region, with extent characterized by a localization length $l_{\text{bd}}\ll L$, in turn attributed to the finite bulk gap in the thermodynamic limit.

\begin{figure}
\includegraphics[scale=0.25]{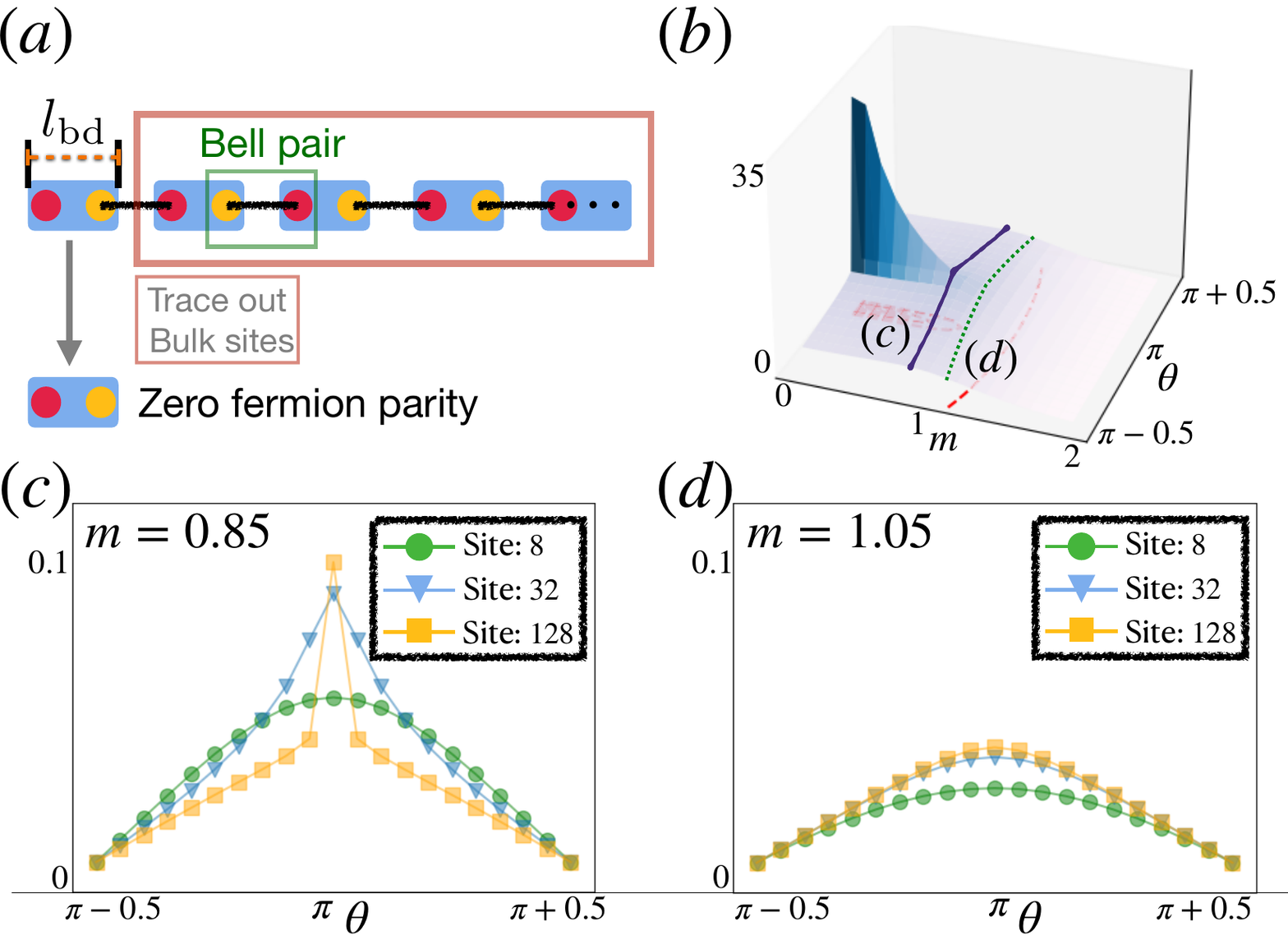}

\caption{Illustration of the bulk-boundary correspondence in an open SSH chain in the dimer limit within the topological regime. In panel (a), $l_{\text{bd}}$ denotes the localization length of the boundary state. The single bond represents the Bell pair. Thus, after tracing over the bulk sites, we are left with a density matrix with zero fermion parity. Alternatively, this zero boundary fermion parity can be inferred via anomaly matching. For the left-most yellow particle, its partition function obtained from tracing out the sites to its right, and equals to $\frac{1}{2}[e^{-i\frac{1}{2}a_0(x_L)}+e^{i\frac{1}{2} a_0(x_L)}]$, which is particle-hole symmetric but picks up a minus sign under the large $U(1)$ transformation. To preserve the underlying on-site large $U(1)$ invariance, the partition function for the left-most red particle must change sign under the large $U(1)$ transformation. Similarly, the partition function for the left-most red particle should be particle-hole symmetric, inferred from the particle-hole symmetry of the density matrix associated with the left-most yellow particle. Together, the large $U(1)$ anomaly and the particle-hole symmetry imply a vanishing fermion parity for the density matrix of the left-most red particle. This analysis is confirmed by the exact calculation of its partition function, i.e., $\frac{1}{2}[e^{-i\frac{1}{2}a_0(x_L)}+e^{i\frac{1}{2} a_0(x_L)}]$. In panel (b), we present numerical results for boundary fermion parity in the $m-\theta$ plane (i.e., $-\log_{10}|\langle e^{-i \theta\hat{Q}_{\text{bd}}}\rangle_{\text{OBC}}|$ with $\theta=\pi$), in a $32$-site open SSH chain at $\beta=2$, where the boundary charge operator has support around the left boundary and spans half the system size. This exhibits cusps in the topological regime along the $\theta$-axis. In panels (c-d), we illustrate the system size dependence of the cusp signal in the topological regime (c) and normal regime (d). Here, we plot $-\log_{10}|\langle e^{-i \theta\hat{Q}_{\text{bd}}}\rangle_{\text{OBC}}|/N$ relative to its minimum value across all $\theta$ values.
\label{supp_fig:ssh}}
\end{figure}

Within this setup, we now demonstrate Eq.~\eqref{supp_eq:fp_bd_charge} via anomaly matching. Specifically, we find that, for odd $\text{ch}_W$, the boundary fermion parity $(-1)^{\hat{Q}_{\text{bd}}}$ supported in $x\in[x_L, x_L+l_{\text{bd}}]$ vanishes, formulated as
\begin{equation}
Z_{\text{bd}}[a_0 = a_0^{\text{bd}}] = 0,\label{supp_eq:boundary_fp_anomaly}
\end{equation}
where $a_0^{{\text{bd}}}(x)$ is a step function equal $\pi$ within $x\in\left[x_{L},\ x_{L}+l_{\text{bd}}\right]$, and $0$ otherwise, and thus Eq. \eqref{supp_eq:fp_bd_charge} follows from Eq.~\eqref{supp_eq:boundary_Z}. This result, $Z_{\text{bd}}[a_0 = a_0^{\text{bd}}] = 0$, is rooted in symmetry principles: (a) The full partition function $Z^{(o)}[a_0]$ retains both particle-hole and large $U(1)$ symmetries; (b) The ``bulk'' partition function $Z^{(p)}[a_0]$, characterized by $S_{\text{eff}}$, maintains only particle-hole symmetry, but exhibits an anomaly under large $U(1)$ transformation, i.e., changing by $(-1)^{\text{ch}_W}$ (as will be shown below); (c) To preserve large $U(1)$ symmetry, the large $U(1)$ anomaly in the bulk $Z^{(p)}[a_0]$ must be absorbed by the boundary $Z_{\text{bd}}[a_0]$, known as anomaly matching.  Together, this leads to the following behavior of the ``boundary'' partition function $Z_{\text{bd}}[a_0]$ under symmetry transformations, 
\begin{equation}
\begin{cases}
\text{Particle-hole symmetry:} & Z_{\text{bd}}\left[a_{0}\right]=Z_{\text{bd}}\left[-a_{0}\right]\\
\text{Large\ \ensuremath{U\left(1\right)}\ anomaly:} & \frac{Z_{\text{bd}}\left[a_{0}+2 a_0^{\text{bd}}\right]}{Z_{\text{bd}}\left[a_{0}\right]}=\left(-1\right)^{\text{ch}_{W}}
\end{cases},\label{supp_eq:anomaly_ph_symmetry}
\end{equation}
which implies, for odd $\text{ch}_{W}$,
\begin{equation}
Z_{\text{bd}}[a_0 = a_0^{\text{bd}}] = 0.\label{supp_eq:boundary_fp_anomaly}
\end{equation} 
Here, the particle-hole symmetry is evident, inherited from the particle-hole symmetry of $Z^{(o)}[a_0]$ and $Z^{(p)}[a_0]$. The large $U(1)$ anomaly condition is derived by smearing out $a_0$ such that it equals $\pi$ within $[x_L,\ x_L + l_{\text{bd}}]$, and then slowly decreases to zero in the bulk. The tail is irrelevant for $Z_{\text{bd}}[a_0]$ due to its boundary nature (cf. Eq.~\eqref{supp_eq:boundary_Z}). In turn, for slowly varying $a_0$, the phase factor of $Z^{(p)}[a_0]$ is captured by $S_{\text{eff}}[a_0]$ obtained in gradient expansion (Eq.~\eqref{supp_eq:S_eff}). Then, integrating over the spatial coordinate $x$ yields
\begin{equation}
    S_{\text{eff}}[a_0] = -\pi \times \text{ch}_W \times \frac{1}{2\pi} \mathcal{I}_W[a_0(x_L)],
\end{equation}
disregarding the $x_R$ term as $a_0(x_R)=0$. This expression highlights a large $U(1)$ anomaly when $a_0(x_L)$ is shifted by $2\pi$, resulting in a change of $S_{\text{eff}}[a_0]\to S_{\text{eff}}[a_0] - \pi \text{ch}_W$, and thus  $Z^{(p)}\rightarrow (-1)^{\text{ch}_W}Z^{(p)}$. The anomaly is compensated by $Z_{\text{bd}}[a_0]$ due to the large $U(1)$ invariance locally at every point in space. Consequently, we obtain Eq.~\eqref{supp_eq:anomaly_ph_symmetry}, from which we infer Eq.~\eqref{supp_eq:boundary_fp_anomaly} using 
\begin{equation}
Z_{\text{bd}}[a_0^{\text{bd}}]=Z_{\text{bd}}[-a_0^{\text{bd}}] = (-1)^{\text{ch}_W}Z_{\text{bd}}[a_0^{\text{bd}}].
\end{equation}
The first equality follows from the particle-hole symmetry in Eq.~\eqref{supp_eq:anomaly_ph_symmetry}, while for the second we use the large $U(1)$ anomaly.
These general results on bulk-boundary correspondence are illustrated in Fig.~\ref{supp_fig:ssh} (a) for the finite-temperature SSH chain (no interactions) in the dimer limit within the topological regime.

In practice, we take a cusp around $\theta=\pi$ in $-\log_{10}|\langle e^{-i \theta \hat{Q}_{\text{bd}}}\rangle_{\text{OBC}}|$ as numerical evidence for the bulk-boundary correspondence.  We take $\hat{Q}_{\text{bd}}$ to have support on half the system, which minimizes finite size effects and provide a necessary condition for the bulk-boundary correspondence. This is because our reasoning above for Eq.~\eqref{supp_eq:fp_bd_charge} implies that for odd $\text{ch}_W$, 
\begin{equation}
\frac{\langle  (-1)^{\hat{Q}_{\text{bd}}^{(l_0)}}\rangle_{\text{OBC}}}{|\langle (-1)^{\hat{Q}}\rangle_{\text{PBC}}|^{l_0/L}} =0\ \ \text{if}\ \ l_0\geq l_{\text{bd}},\label{supp_eq_bdfp}
\end{equation}
with $\hat{Q}^{(l_0)}_{\text{bd}}$ for the boundary charge operator supported in $[x_L, x_L +l_0]$, which originates from the localized nature of $Z_{\text{bd}}[a_0]$: specifically, a $\pi$-valued bump in $a_0$ over a length larger than $l_{\text{bd}}$ at the boundary causes the $Z_{\text{bd}}[a_0]$ to vanish. 

A numerical illustration is provided in Fig.~\ref{supp_fig:ssh}  for the free SSH chain. Results for the interacting problem are given in the main text, with additional data presented in the following subsection.

\begin{figure}
\includegraphics[scale=0.25]{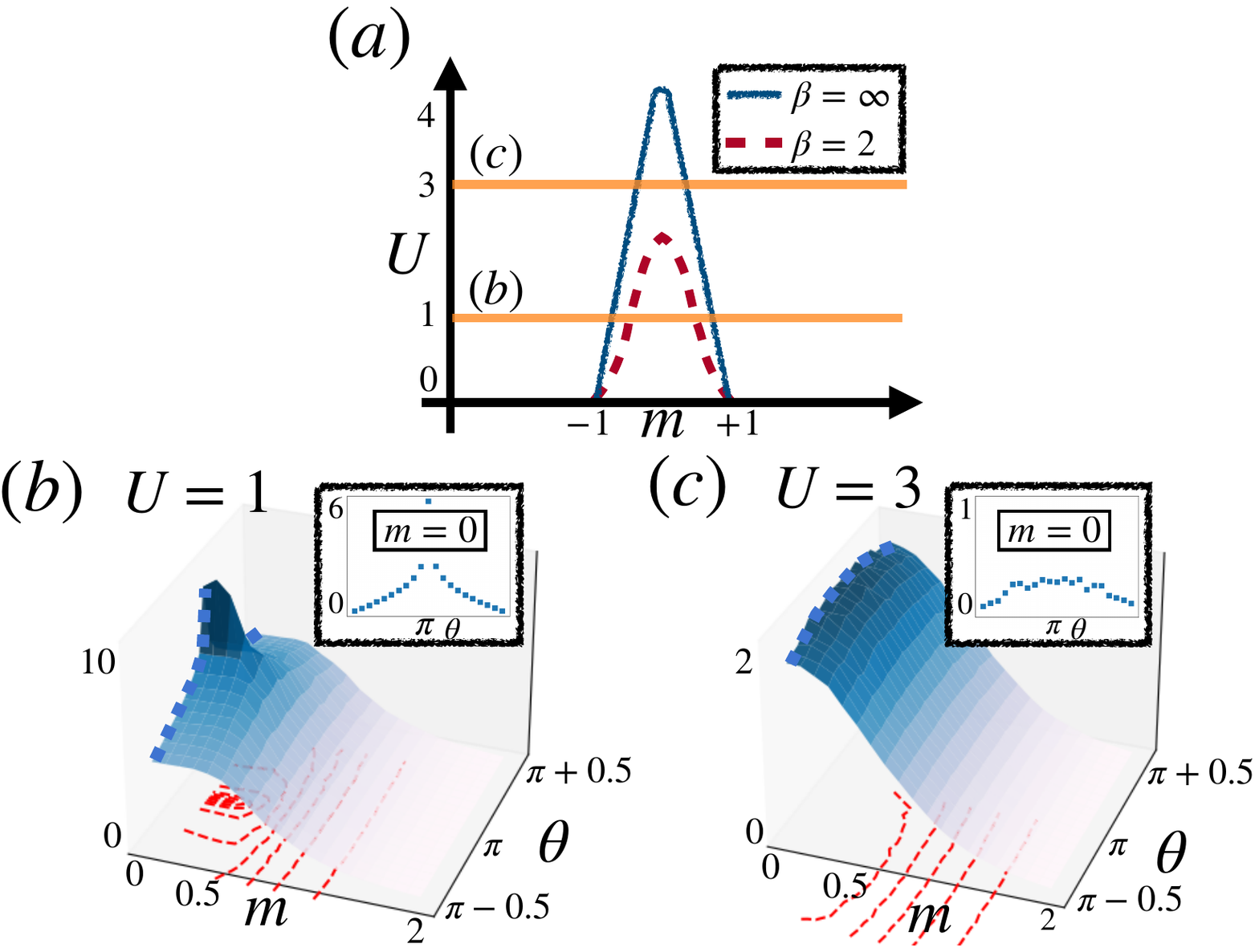}

\caption{Numerical results for the boundary fermion parity (i.e., $-\log_{10}|\langle e^{-i \theta \hat{Q}_{\text{bd}}}\rangle_{\text{OBC}}|$ with $\theta=\pi$) as a function of $m$ at $U=1$ (b) and $U=3$ (c). We analyze two cuts in the phase diagram of a 32-site \textit{open} SSH-Hubbard chain with $\beta=2$, depicted in panel (a) (red dashed line).  The boundary charge operator $\hat{Q}_{\text{bd}}$ is supported around the left boundary and spans 16 sites.  (b) Within the topological regime, the boundary fermion parity exhibits a cusp and takes small values (i.e., large in terms of $-\log_{10}|\langle (-1)^{\hat{Q}_{\text{bd}}}\rangle_{\text{OBC}}|$). (c) In the normal regime,  it behaves smoothly  (and remains small in terms of $-\log_{10}|\langle (-1)^{\hat{Q}_{\text{bd}}}\rangle_{\text{OBC}}|$).
\label{supp_fig:ssh_hubbard_edge}}
\end{figure}

\begin{figure}[t!]
\includegraphics[scale=0.25]{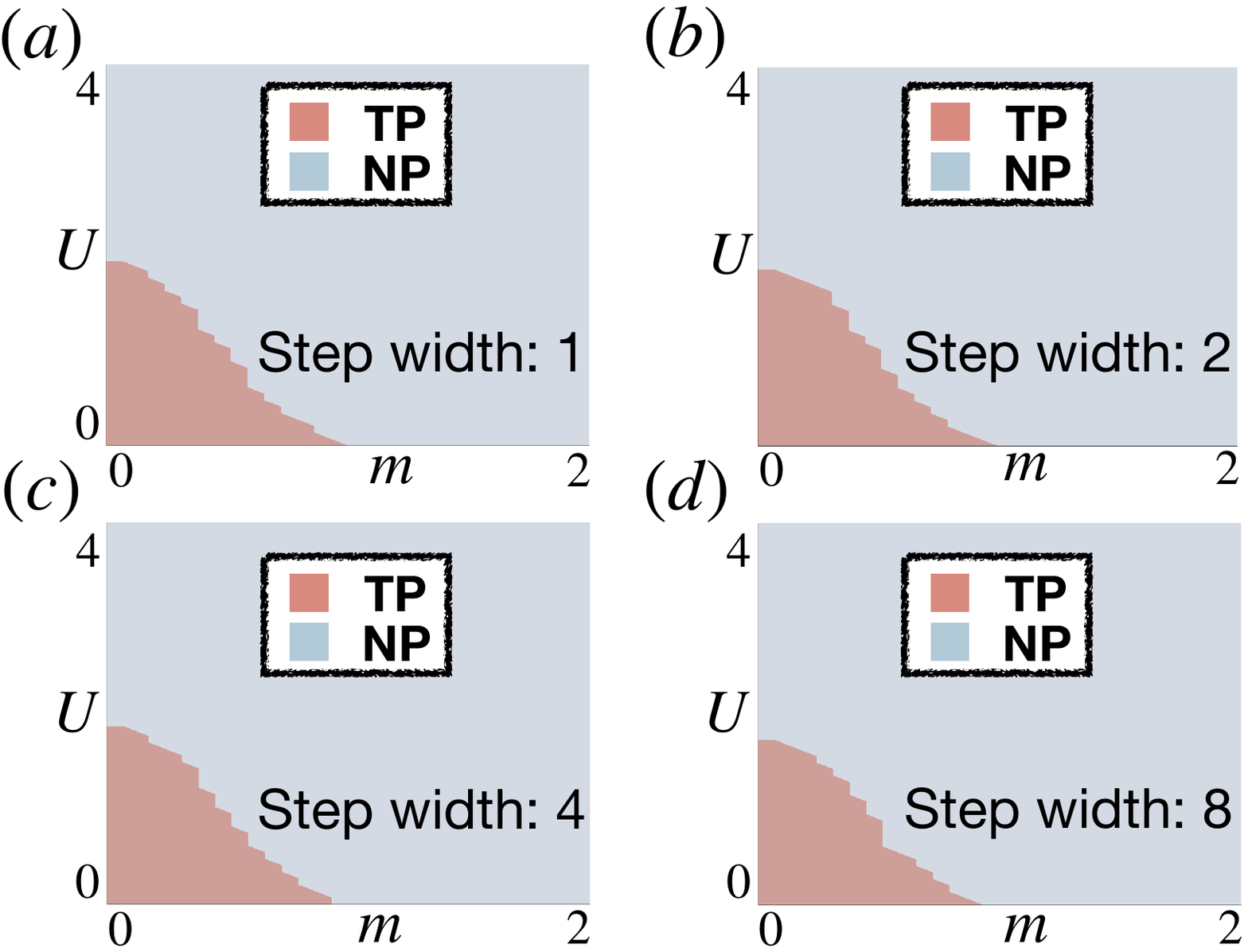}

\caption{Numerical results for the phase diagram in a $32$-site SSH chain at $\beta=2$, obtained from different $a_0(x)$ profiles ($S_{\text{eff}} =0 (\pi) $ in the normal (topological) regimes): (a) a linear profile, i.e., $a_0(x)=\frac{2\pi}{L}x$ (step width of one lattice constant, denoted by $d_{\text{lat}}$ and set to $1$); (b) a profile with step width $2$, i.e., $a_0(x)=\frac{2\pi}{L}(1, 1, 3, 3, \dots)$; (c) a profile with step width $4$; and (d) a profile with step width $8$. 
\label{supp_fig:ssh_a0}}
\end{figure}
\subsection{Numerical results \label{supp_sec:bbc_numerical}}

Here we first provide numerical evidence for the boundary fermion parity, in order to test the bulk-boundary correspondence. Specifically, Fig. \ref{supp_fig:ssh_hubbard_edge} focuses on the boundary fermion parity, cf. panels (b-c). It behaves consistently with the finite-temperature phase diagram shown in panel (a), extracted via the bulk topological order parameter.

Second, we complement and test numerically the effective field theory approach, according to which the critical line should not depend on the precise choice of the slowly varying function $a_0$, as long as the symmetries are respected. To this end, we compute the phase factor $S_{\text{eff}}$ for various $a_0$ profiles, including step functions with widths of 1, 2, 4, and 8 lattice constants ($d_{\text{lat}}=1$). This yields consistent phase diagrams see Fig.~\ref{supp_fig:ssh_a0}. Small deviations start to be visible only for step size 8, which we attribute to finite size effects enhanced by large steps in the profile function $a_0$.

\section{Further aspects of the phase diagram}

\begin{figure}[t]
\includegraphics[scale=0.25]{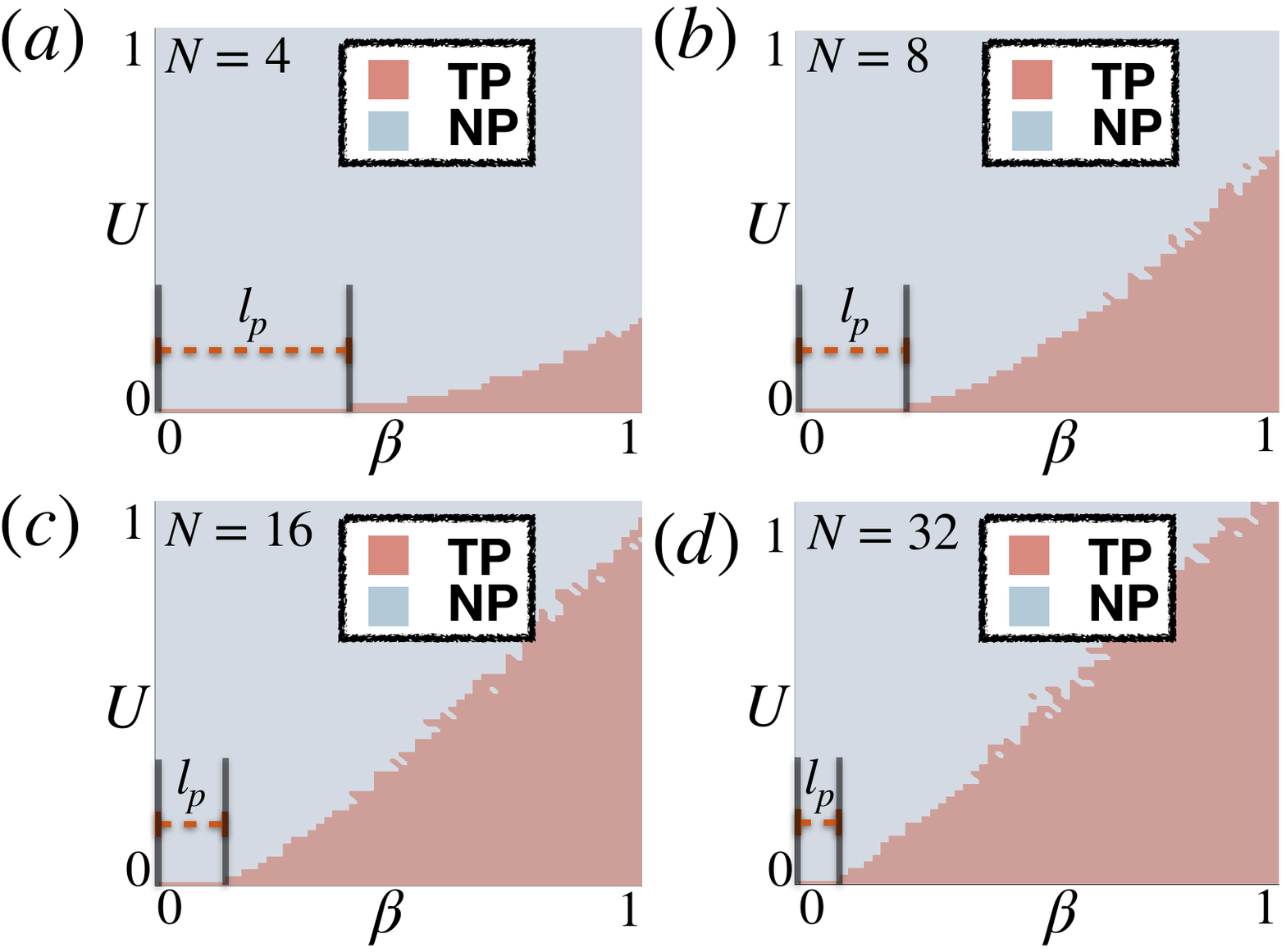}
\includegraphics[scale=0.13]{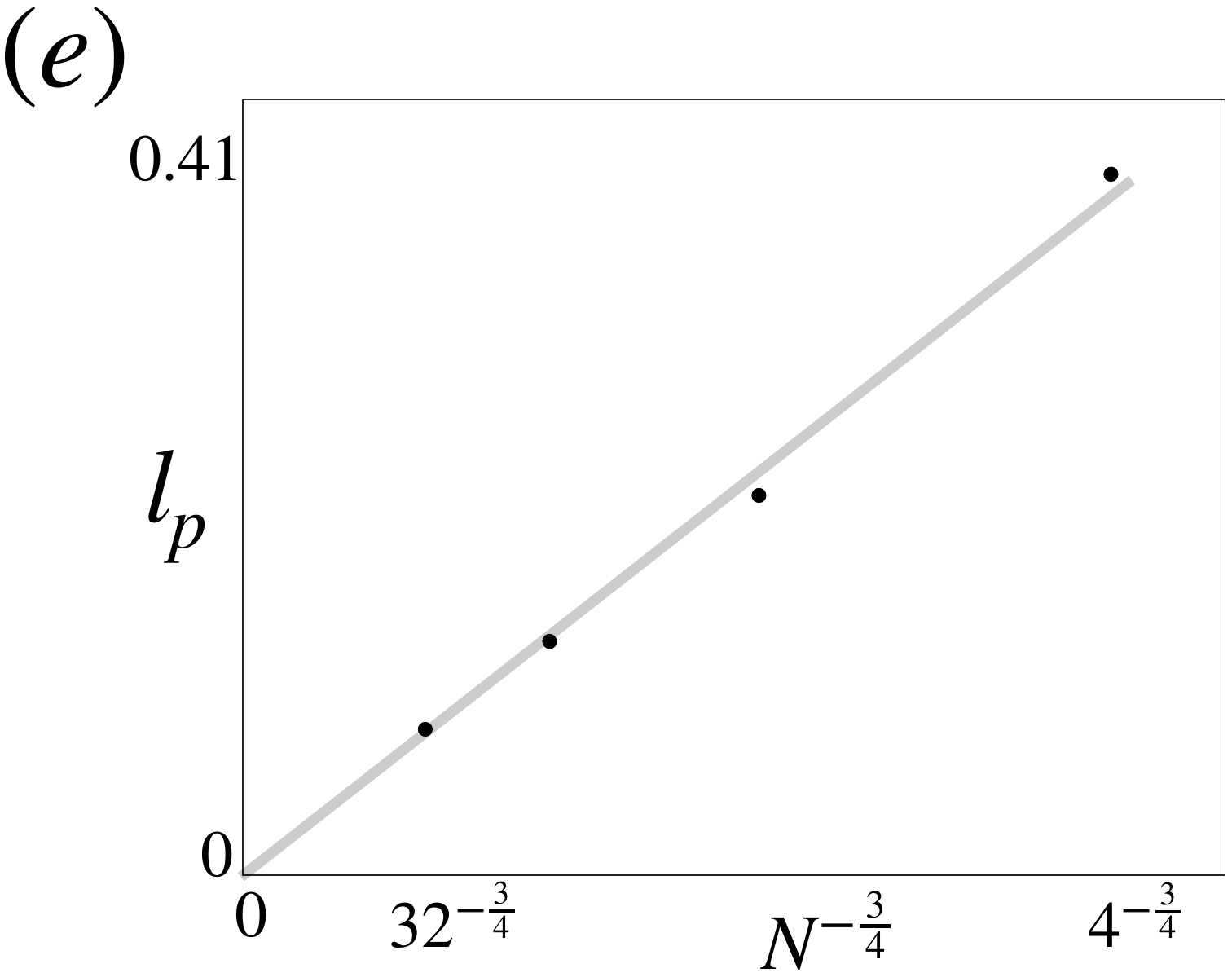}

\caption{Plot of the phase diagram of the SSH-Hubbard model in the $U-\beta$
plane for different system site numbers ($N=L/d_{\text{lat}}$) with $N=4,\ 8,\ 16,\ 32$ in
(a-d). This phase diagram is obtained via the DQMC averaged over $500$ samples.
These plots exhibit a phase boundary with a flat profile at high temperature
with a plateau width $l_{p}$, subsequently transitioning to a linear
one (i.e., $\beta_{c}\propto U$ and $\beta_{c}^{-1}$ for the critical
temperature). Notably, the plateau width $l_{p}$ shrinks to zero with increasing site number, see (e) for a finite-size scaling plot.
\label{supp_fig:U_beta_phase_finitesize}}
\end{figure}

We provide additional numerical results for the phase diagram
of the SSH-Hubbard model in the $U-\beta$ plane for $m=0$, with focus
on its system size dependence, Fig. \ref{supp_fig:U_beta_phase_finitesize}. The results suggest the scaling of the inverse critical temperature $\beta_c \propto U$ at high temperatures in the thermodynamic limit.

\begin{thebibliography}{61}
\bibitem [{FN_()}]{FN_dw}%
  \BibitemOpen
  \href@noop {} {}\bibinfo {note} {We focus on domain wall configurations that
  interpolate between topologically distinct phases, as other configurations do
  not impact the topological signal, as we confirm numerically.}\BibitemShut
  {Stop}%
\bibitem [{fn_()}]{fn_ph2}%
  \BibitemOpen
  \href@noop {} {}\bibinfo {note} {A particle-hole pair with energies
  $\left\{\mathcal{E}, -\mathcal{E}\right\}$ yields a contribution
  $2[1+\cosh(\beta\mathcal{E})]$, which is a monotonically increasing function
  of $\mathcal{E}$.}\BibitemShut {Stop}%
\bibitem [{FN_()}]{FN_Iw}%
  \BibitemOpen
  \href@noop {} {}\bibinfo {note} {$\mathcal{I}_W|_{a_0}^{a_0+2\pi}$ is defined
  modulo $2\pi$, with the integer part absorbed into
  $\text{ch}_W$.}\BibitemShut {Stop}%
\end{thebibliography}
\end{document}